# Experimental Constraints of Using Slow-Light in Sodium Vapor for Light-Drag Enhanced Relative Rotation Sensing

**Renu Tripathi, G.S. Pati, M. Messall, K. Salit and M.S. Shahriar**

*Department of Electrical and Computer Engineering,*

*Northwestern University, Evanston IL 60208*

***PACS Codes:*** 45.40.Cc, *42.50.-p, 42.50.Nn,* 82.70.-y, 39.20.+q, 39.90.+d

***Abstract:***

We report on experimental observation of electromagnetically induced transparency and slow-light ($v_g \approx c/607$) in atomic sodium vapor, as a potential medium for a recently proposed experiment on slow-light enhanced relative rotation sensing [11]. We have performed an interferometric measurement of the index variation associated with a two-photon resonance to estimate the dispersion characteristics of the medium that is relevant to the slow-light based rotation sensing scheme. We also show that the presence of counter-propagating pump beams in an optical Sagnac loop produces a backward optical phase conjugation beam that can generate spurious signals, which may complicate the measurement of small rotations in the slow-light enhanced gyroscope. We identify techniques for overcoming this constraint.

***Key Words*:** Rotation sensing, optical gyroscope, slow light, electromagnetically induced transparency, sodium vapor, Mach-Zehnder interferometer

Extreme dispersion induced by electromagnetically induced transparency (EIT) can reduce the speed or group velocity of light by many orders of magnitude compared to the speed of light in vacuum [1-4]. Recently, there has been a significant interest in the physics and applications of slow light. Typical applications include schemes where a controllably varied group velocity is used to realize optical delay lines, buffers, etc. [5, 6], as well as techniques where reversible mapping of photon pulses in atomic medium are used for quantum state storage [7-9]. Recent proposals have also envisioned using slow light to enhance the rotational sensitivity of an interferometric optical gyroscope [10, 11]. Such an interferometer may use slow light induced dispersive drag for enhanced sensitivity in relative rotation sensing. In this case, the rotational fringe shift is augmented by the group index or the dispersion in the medium, which, for realistic conditions, can yield many orders of magnitude improvement in the sensitivity of the gyroscope [11].

An experimental implementation of the interferometric gyroscope relies on using an EIT medium, so that the counter-rotating optical fields experience resonant dispersion along the entire optical path. A relative motion between the medium and interferometer is also needed [11]. This gives rise to a rotational fringe shift that depends on the magnitude of the dispersion in the medium. We have considered Na atoms in a dilute vapor as an example of an experimental medium for this purpose. We have studied EIT in Doppler-broadened optical transitions of the $D_1$ line in Na vapor, and experimentally measured its dispersion characteristics that are relevant to its use in a slow-light enhanced Sagnac interferometer. In particular, magnitudes of the index change and the dispersion, under a narrow EIT resonance, have been measured by a phase delay obtained using a homodyne detection scheme. Precise measurements of these values help us infer the dynamic range as well as the magnitude of the sensitivity enhancement [11]. An excellent agreement is obtained while comparing the magnitudes of the first-order dispersion estimated from the time-delay of slowed optical pulses and the slope of the dispersion curve obtained from the interferometric measurements discussed above.

Figure 1 shows energy levels selected for a three-level Λ-type system in the $D_1$ line of Na. A frequency-tuned cw dye laser with a narrow linewidth (< 1 MHz) is used to derive the optical beams. The pump beam is locked to the $3^2S_{1/2}$, F=1 and $3^2P_{1/2}$, F'=1 transition using saturated absorption. The probe laser is derived by frequency shifting the pump beam with an acousto-optic modulator (AOM). The frequency difference between the pump and probe laser field is set equal to the frequency-splitting (1.772 MHz) between the ground states $3^2S_{1/2}$, F=1 and $3^2S_{1/2}$, F=2.

The experimental arrangement is shown in Figure 2. The pump and the probe beams are cross-linearly polarized. They propagate collinearly in a 10 cm long sodium vapor cell. The beams are focused inside the cell to a beam waist (1/e in intensity) of ~100 μm that corresponds to a confocal distance of ~ 5 cm. The orthogonality of the pump and the probe polarizations allows us to filter the pump at the output in order to measure the absorption and dispersion properties of the probe field very accurately. As the laser excitation in alkali atoms like sodium involves many hyperfine Zeeman sublevels, the Λ-type scheme often departs from an ideal EIT system in the presence of a stray magnetic field. The cell is, therefore, magnetically shielded using two-layers of μ-metal to minimize the effects of stray magnetic fields. During the experiment, the sodium cell is heated to a steady temperature of 100$^{o}$C using bifilarly wound coils that produce a negligible magnetic field.

As shown in figure 2, the frequency of the probe is continuously scanned around the two-photon resonance condition, using a double-pass acousto-optic frequency shifter to observe the linewidth of EIT. The pump intensity is set to nearly 20 W/cm$^2$, which corresponds to a Rabi frequency $\Omega$ ~ 40 GHz. The ratio of the pump to probe intensities is set to ~ 10. The EIT linewidth is found to be ~ 1 MHz. This is limited by the transit time (~ 1 μs) of the atoms and can be improved upon by adding a buffer gas in the vapor cell. A maximum probe transmission of 23% has been observed. Figure 3a shows a sequence of EIT resonances with increasing pump intensity. A frequency-dithered lock-in-detection is used to observe the EIT signal in the presence of the residual-pump beam. Figure 3b shows the corresponding change in the slope of the lock-in-detection signal with increasing in pump intensity.

The dispersion characteristic associated with sub-natural EIT line-widths is measured using a homodyne method [14] based on a Mach-Zehnder interferometric configuration, as shown in Figure 2. An unperturbed fraction of the probe beam traversing an equivalent optical path outside the cell is used as a reference beam in the homodyne detection scheme. The reference beam and the transmitted EIT signal are interferometrically combined at the output. The signal intensity detected on the photodiode is proportional to the phase shift $\Delta\phi = (2\pi/\lambda)[n(\omega)-1]L$, introduced by the dispersion of the atomic medium, and is given by $i_D \propto 2|E_p||E_{ref}|\cos[\Delta\phi + \phi_{ref}]$, where $E_p$ and $E_{ref}$ are the amplitudes of the probe and the reference, respectively, L is the active interaction length, and $\phi_{ref}$ is the phase of the reference beam. A piezo-mounted mirror is used to adjust the reference phase to $\pi/2$ such that the observed magnitude of the photo-current is directly proportional to $\Delta n(\omega)$ $[= n(\omega) - 1]$ for $|k\,\Delta n(\omega)\,L| << 1$, valid if $|\Delta n| << 10^{\ 6}$ (typical for a dilute atomic medium). The frequency of the probe laser is swept at a faster rate (1 KHz) so that only a negligible drift occurs between the interferometer arms while measuring the phase delay $\Delta\phi$. Figure 4 shows the index variation as a function of the difference frequency $\delta$. Several measurements were taken to measure the slope of the positive dispersion profile at the center of the EIT resonance i.e. $\left(\partial n/\partial \omega\right)\big|_{\delta=0}$ (~ 1.89 x $10^{-13}$ $rad^{-1}sec$) that corresponds to a slow group velocity $v_g$ (~ c/607) in the medium. This also corresponds to a refractive index variation $\Delta n$ ~ 1.89 x $10^{-7}$ over a probe frequency bandwidth $\Delta f$ = 1 MHz. The accuracy of these measurements is subsequently verified by pulse delay measurements from a slow light experiment. Similarly, the magnitude of the second-order dispersion $\left(\partial^2 n/\partial \omega^2\right)\big|_{\delta=0}$ has also been estimated from a polynomial fit to the dispersion profile, over a frequency range $\Delta f$ = 1 MHz. This value being small (2.24 x $10^{-15}$ $rad^{-2}sec2$) corresponds to negligible pulse spreading during slow pulse propagation in the EIT medium. Our results suggest that the present atomic medium can be employed in an interferometric gyroscope to demonstrate drag-induced sensitivity enhancement nearly by a factor of 600 (equal to group index $n_g$). For improved performance, the slope of the linear dispersion can be increased using a medium with larger transit times by adding a buffer gas into the cell.

In order to observe the slowing of probe pulses due to the steep normal dispersion in the EIT medium, short probe pulses (FWHM ~ 400 ns) with an arbitrary delay and smooth profile (nearly transform limited) are generated using an RF mixer and a digital pulse generator with a frequency filter. During the experiment, the pump beam remains on continuously. The group velocity, $v_g$, of the probe pulse, is estimated accurately by measuring the time delay with respect to a reference probe pulse propagating outside the cell. This is then used to measure indirectly the linear dispersion coefficient $\left(\partial n/\partial \omega\right)\big|_{\delta=0}$ at the center of the EIT resonance, and is compared with the complementary interferometric measurements.

Figure 5 shows the slowed probe pulses with respect to a reference pulse and the corresponding time delays. The presence of residual magnetic fields due to the heating coils provide additional mechanisms for dephasing and thus broaden the EIT linewidth and reduce the dispersion induced time delay. A maximum time delay of ~ 202 ns has been observed by temporarily switching off the current to the coils. The value of $v_g$ corresponding to this time delay is found to be ~ c/607, which agrees very well with our previous measurement. As can be also seen in fig. 5, negligible pulse spreading is observed over the input pulse bandwidth. This also confirms the fact that the dispersion is linear over the pulse bandwidth. Figure 6 shows significant pulse spreading when the carrier frequency of the probe is detuned above the two-photon resonance condition. In this case, the pulse undergoes asymmetric spreading as it experiences dispersion which is no longer linear over one half of its bandwidth. Also, as the frequency is detuned away from resonance, the transmission of the probe pulse decreases and smaller time delays are observed with respect to the reference pulse, as expected.

To realize the feasibility of light drag induced rotational sensitivity enhancement near an atomic resonance, we have considered a common path Sagnac interferometer where the atomic medium, besides the forward-propagating pump and probe, also encounters a back-propagating pump resulting from the common path beam, as shown in fig. 7. This geometry, while crucial in canceling the effect of external vibrations for example, can form a four-wave mixing (FWM) process and write nonlinear gratings via EIT to generate a backward optical phase conjugate (PC) beam. In order to test this effect, we used a back-propagating pump which is frequency degenerate and has the same polarization as the forward pump [12]. The process can be understood in terms of a two-photon induced grating formed in the ground-state coherence $\rho_{12} \propto \Omega' \Omega_b^* \big/ \left(\Omega'^2 + \Omega_b^2\right)^{1/2} \exp\left\{-i\left[(\omega_2 - \omega_1)t - (k_2 - k_b)z\right]\right\}$ by the probe and backward pump, when the atoms are optically pumped into the dark superposition state. Here $\Omega_b$, $\Omega'$ are the Rabi frequencies, $k_b$, $k_2$ are propagation vectors and $\omega_1$, $\omega_2$ are optical frequencies associated with the backward pump and probe, respectively. The forward pump produces a phase matched read-out of the grating to generate a backward PC beam at the probe frequency. It is observed experimentally that even though the forward pump atomic transition is Doppler broadened, only a low intensity backward pump is needed to produce the PC beam. Figure 8 shows the measured phase conjugate reflectivity at the beam splitter output (figure 2) as a function of the probe detuning. The frequency width (FWHM ~ 1 MHz) is nearly equal to the EIT linewidth and the measured PC reflectivity is ~ 0.017. Such a signal can circulate in an optical Sagnac loop and serve as a source of backscattering, giving rise to discrepancies in fringe shift measurements due to small rotations in a sensitive slow light based gyroscope. The problem may be circumvented either in a common path geometry using an active medium partially filling the arms of the interferometer, or by using a Mach-Zehnder type Sagnac interferometer whose arms are entirely filled with the active medium. Here, one can prevent the back-propagating pump from traversing the entire path by using polarization selective elements in the interferometer. However, in a non-common path configuration, additional stability to fringe shift due to optical path length variation has to be provided by an external feedback and locking mechanism.

In conclusion, we have experimentally investigated the resonant dispersion characteristic of a Na atomic medium as a possible candidate for a light-drag enhanced relative rotation sensor. We have determined that an enhanced factor of ~ 600 is readily achievable. Such a medium is currently being used in an optical Sagnac interferometer to observe sensitivity enhancement. We have also shown that the presence of an unavoidable back-propagating pump in a common path interferometer produces optical phase conjugation at low optical power, which is detrimental to fringe shift measurement, and have suggested schemes that can help circumvent this problem.

***Acknowledgements:***

This work was supported in part by the AFOSR and the ARO MURI program.

**List of Figures:**

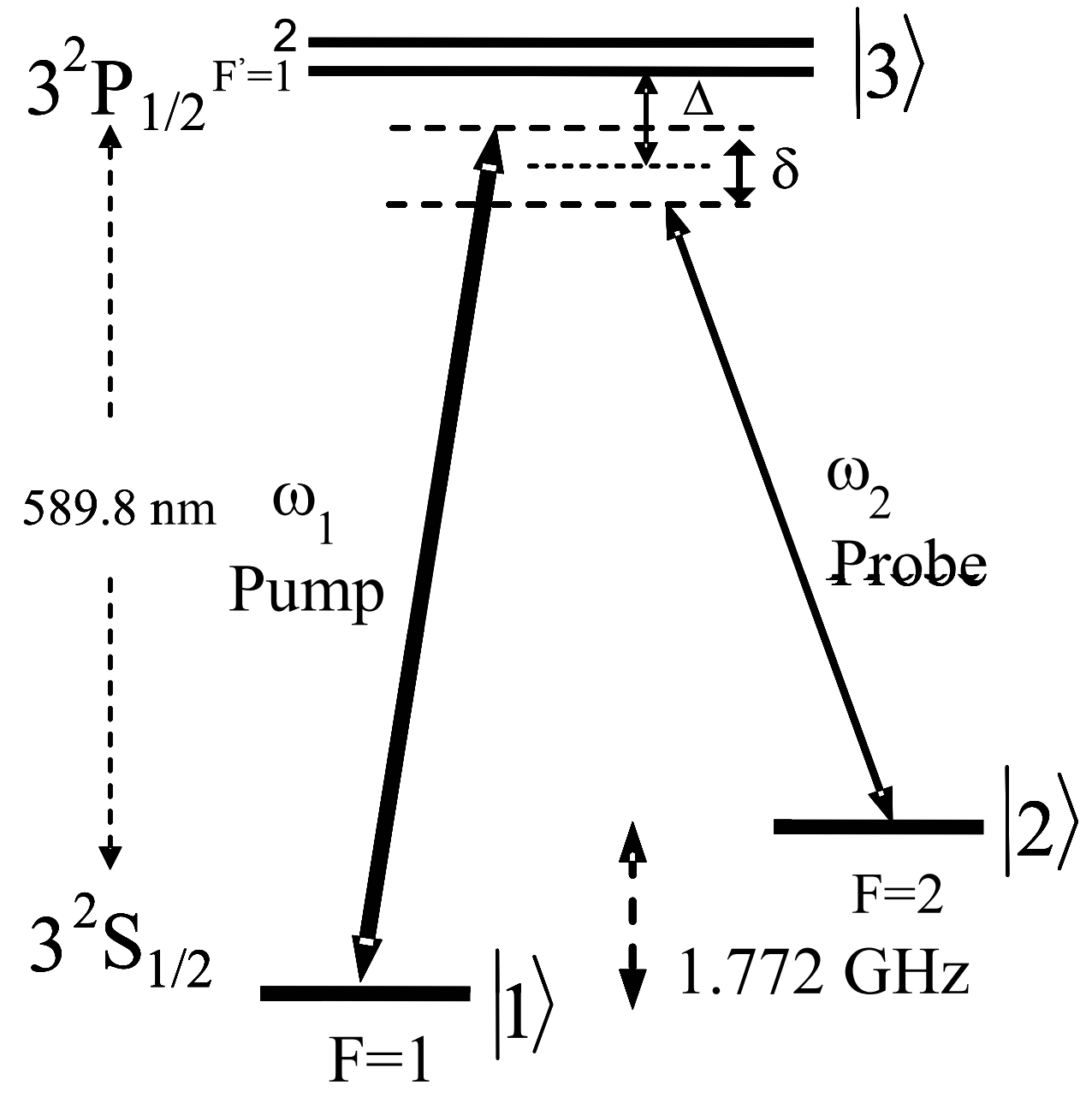

Fig. 1 Energy level structures in $D_1$ line of Sodium used as a three-level Λ-system

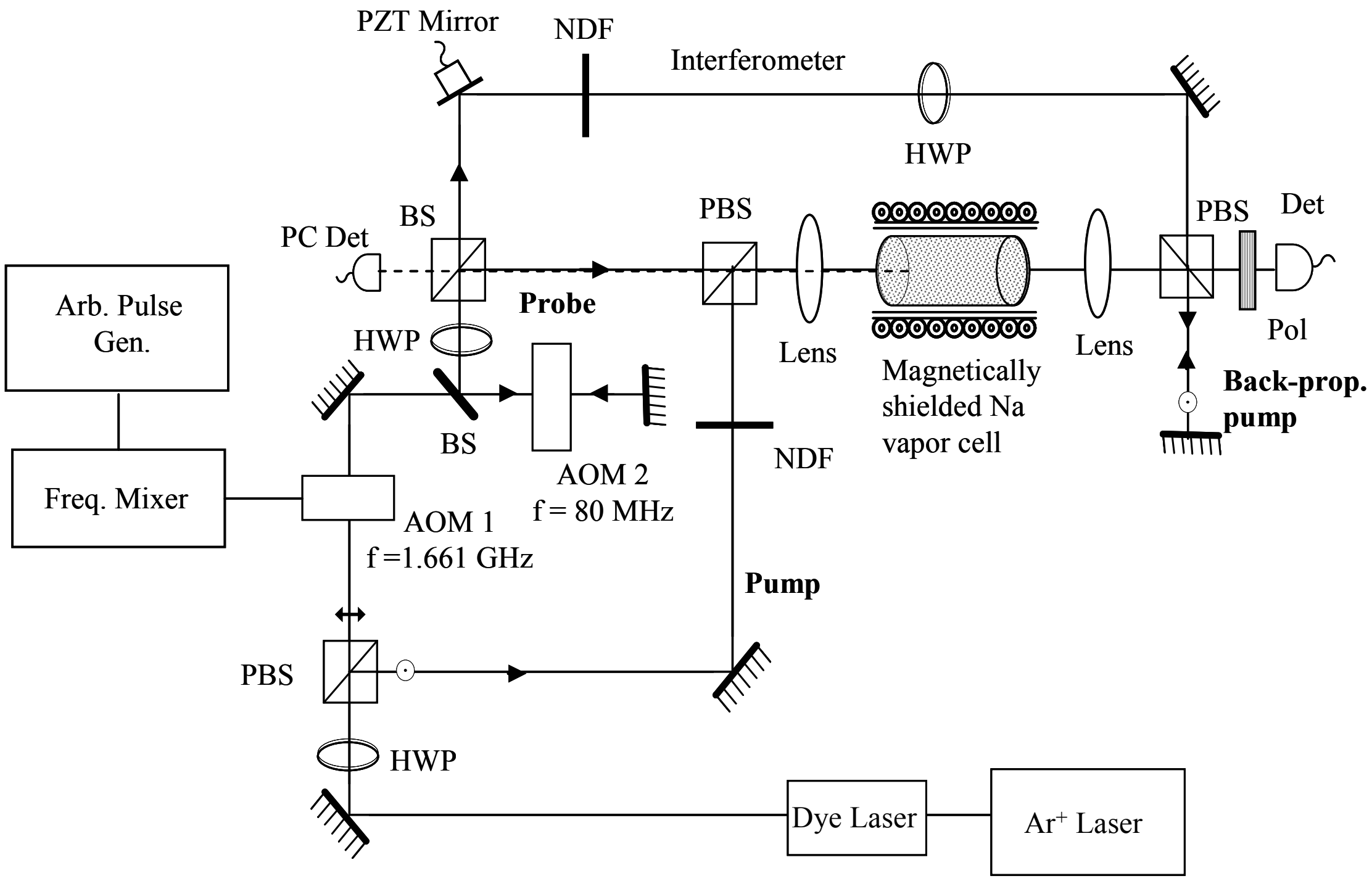

Fig. 2 Experimental setup used to observe EIT, slow light and PC in Na vapor. HWP, half-wave plate; PBS, polarizing beam splitter; NDF, neutral density filter; AOM, acousto-optic modulator; Pol, polarizer; Det, detector.

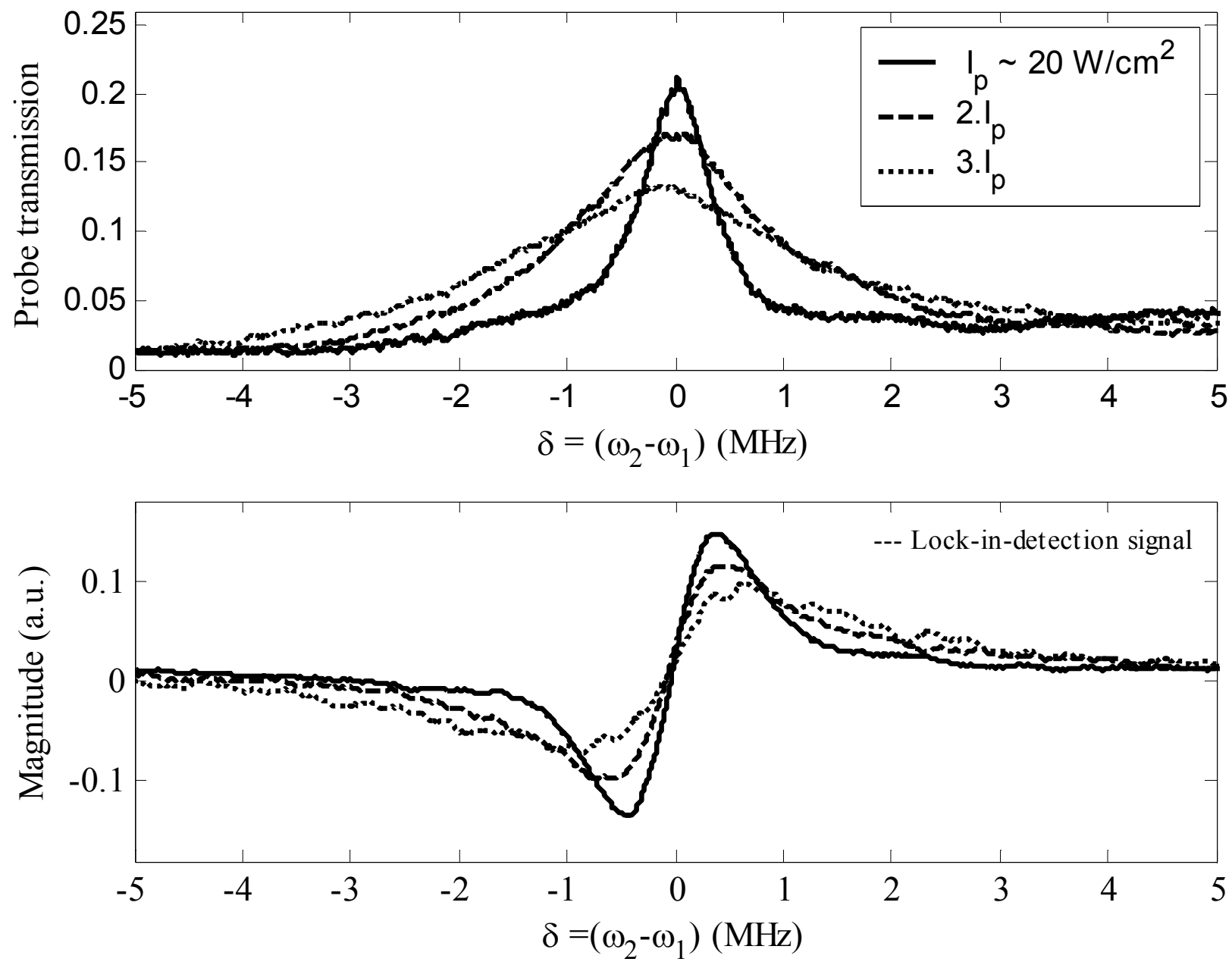

Fig. 3 Variation of EIT line width with pump intensity (a) detector output (b) lock-in-detection signal.

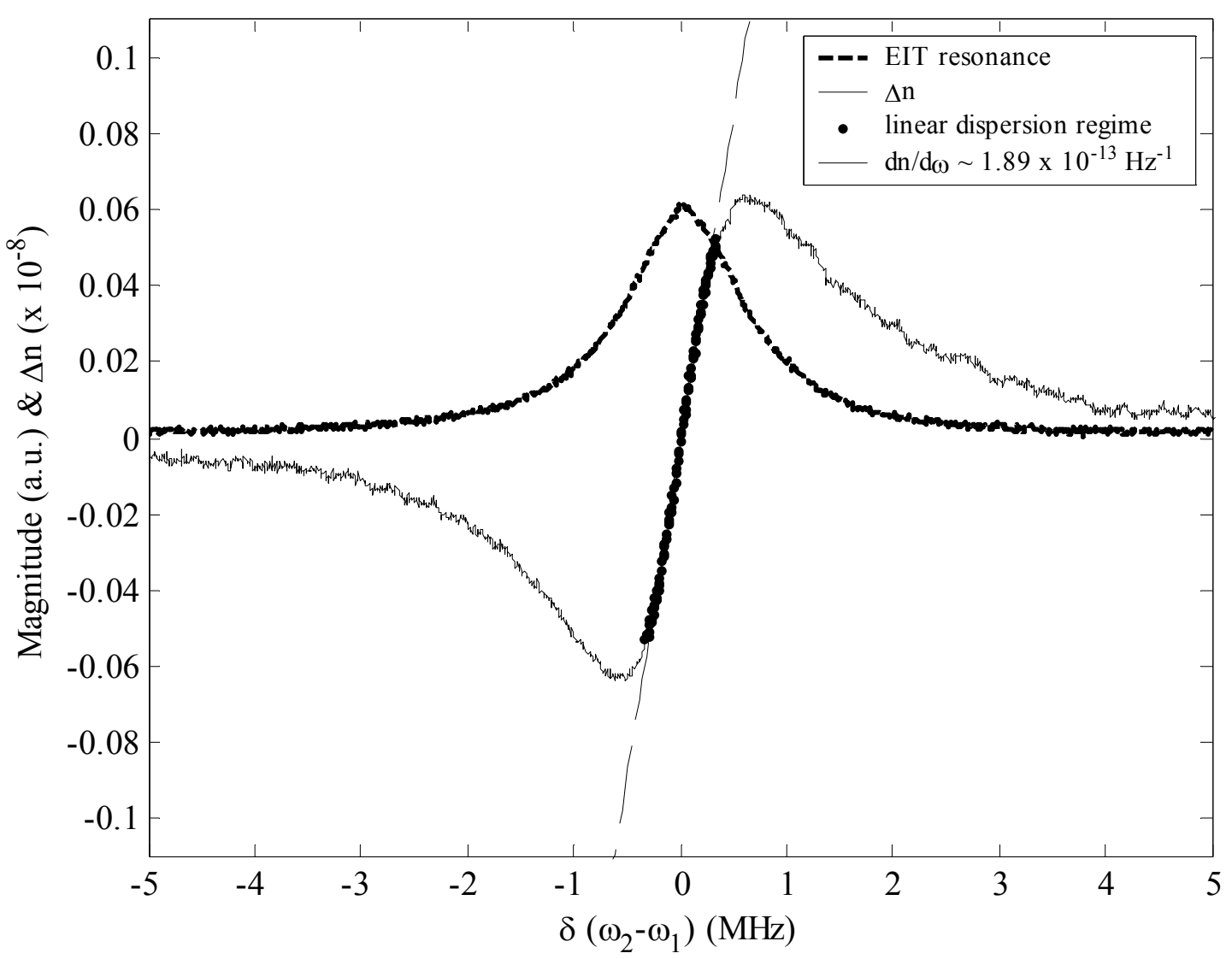

Fig. 4 Interferometrically measured refractive index variation associated with EIT dispersion

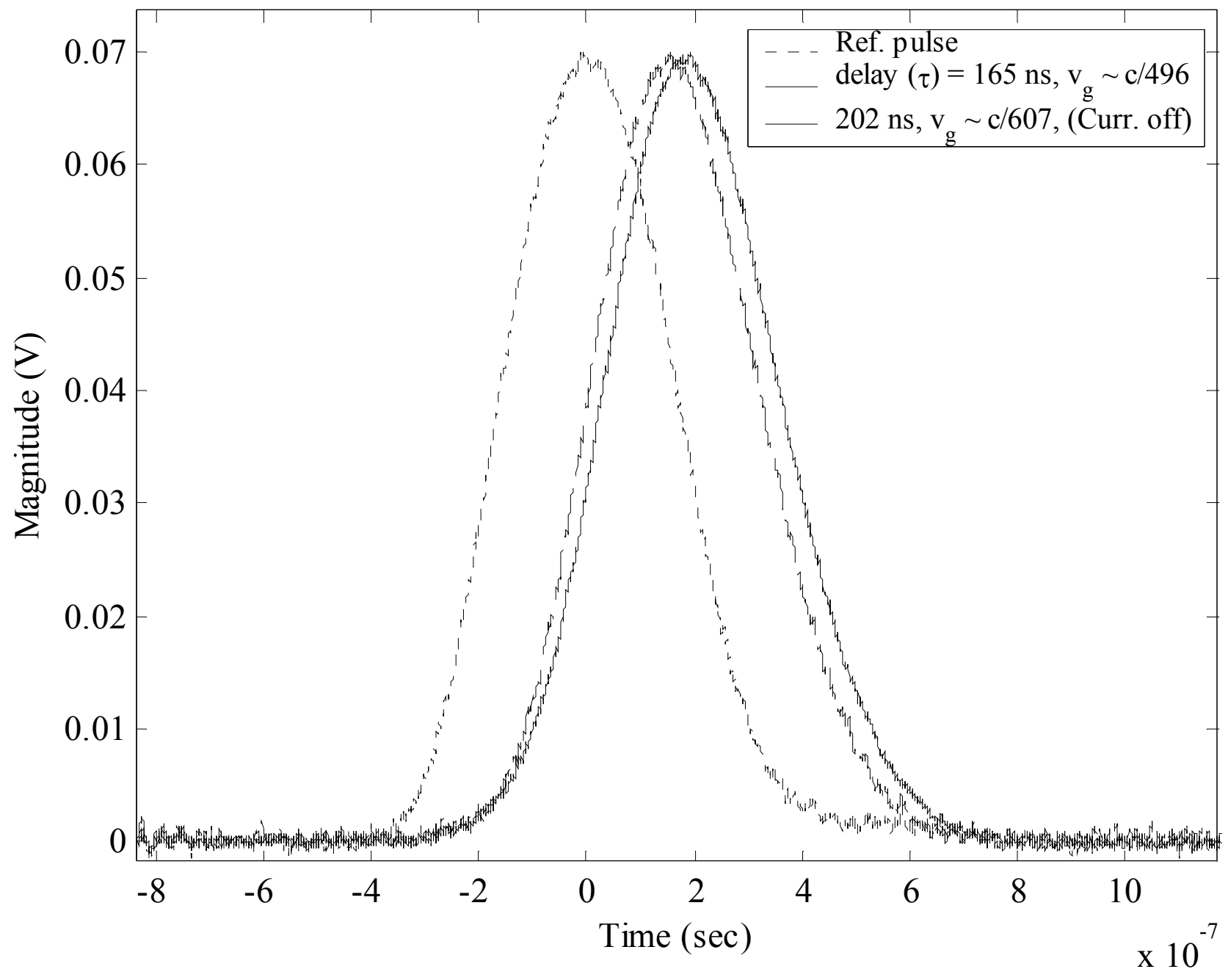

Fig 5 Probe pulse slowing using EIT induced dispersion in sodium vapor

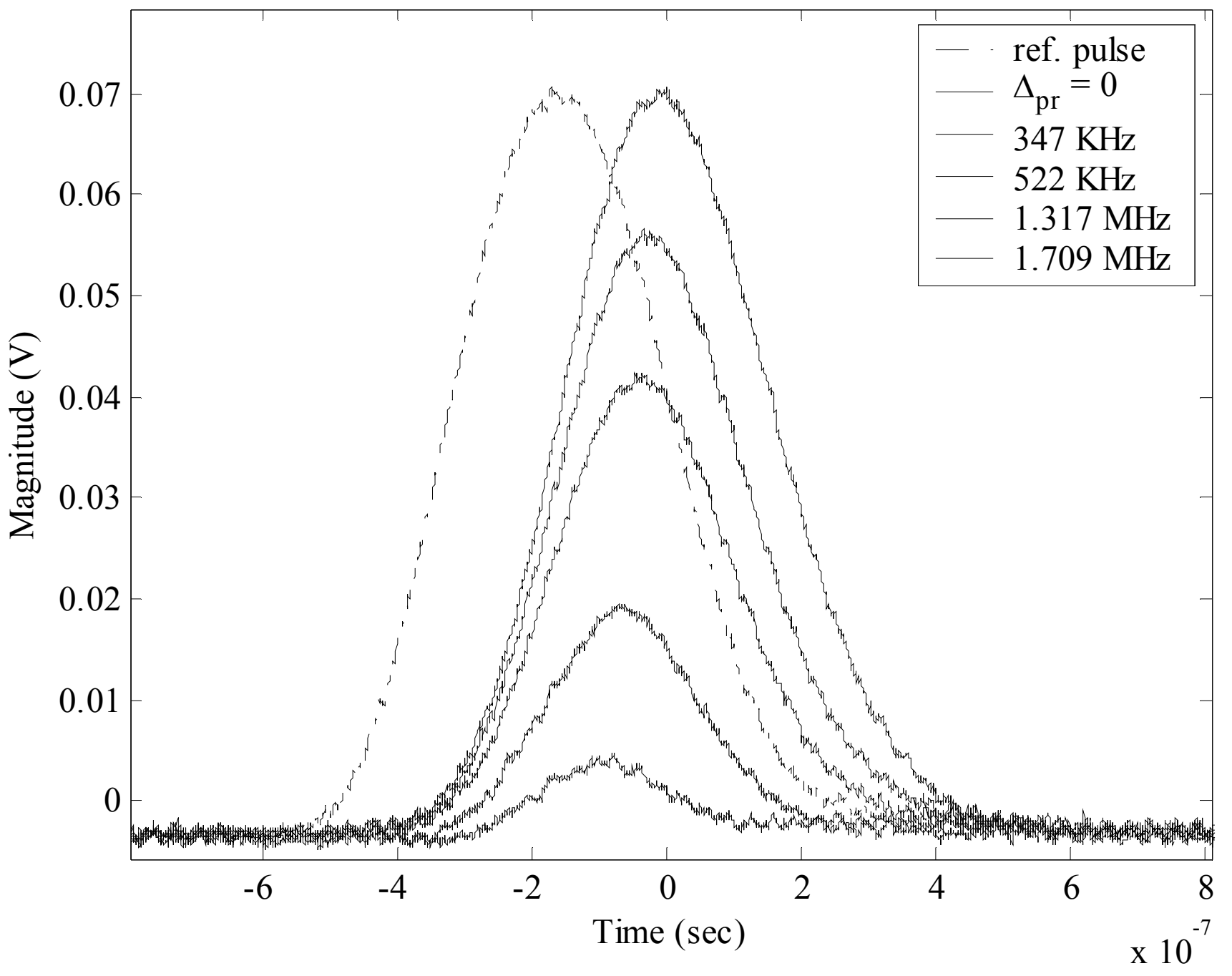

Fig. 6 Pulse spreading due to second-order dispersion observed by frequency detuning the probe carrier frequency away from two-photon resonance

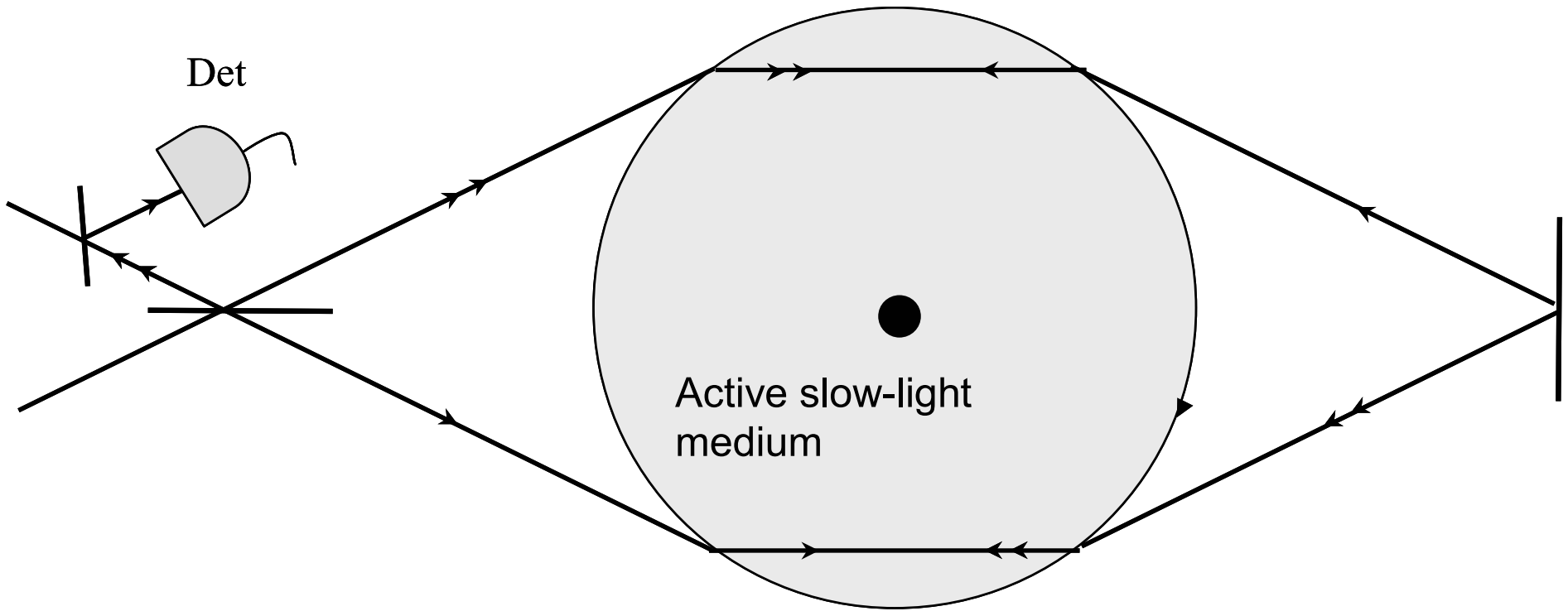

Fig. 7 A common path Sagnac interferometer containing a slow-light medium for light-drag enhanced rotation sensing

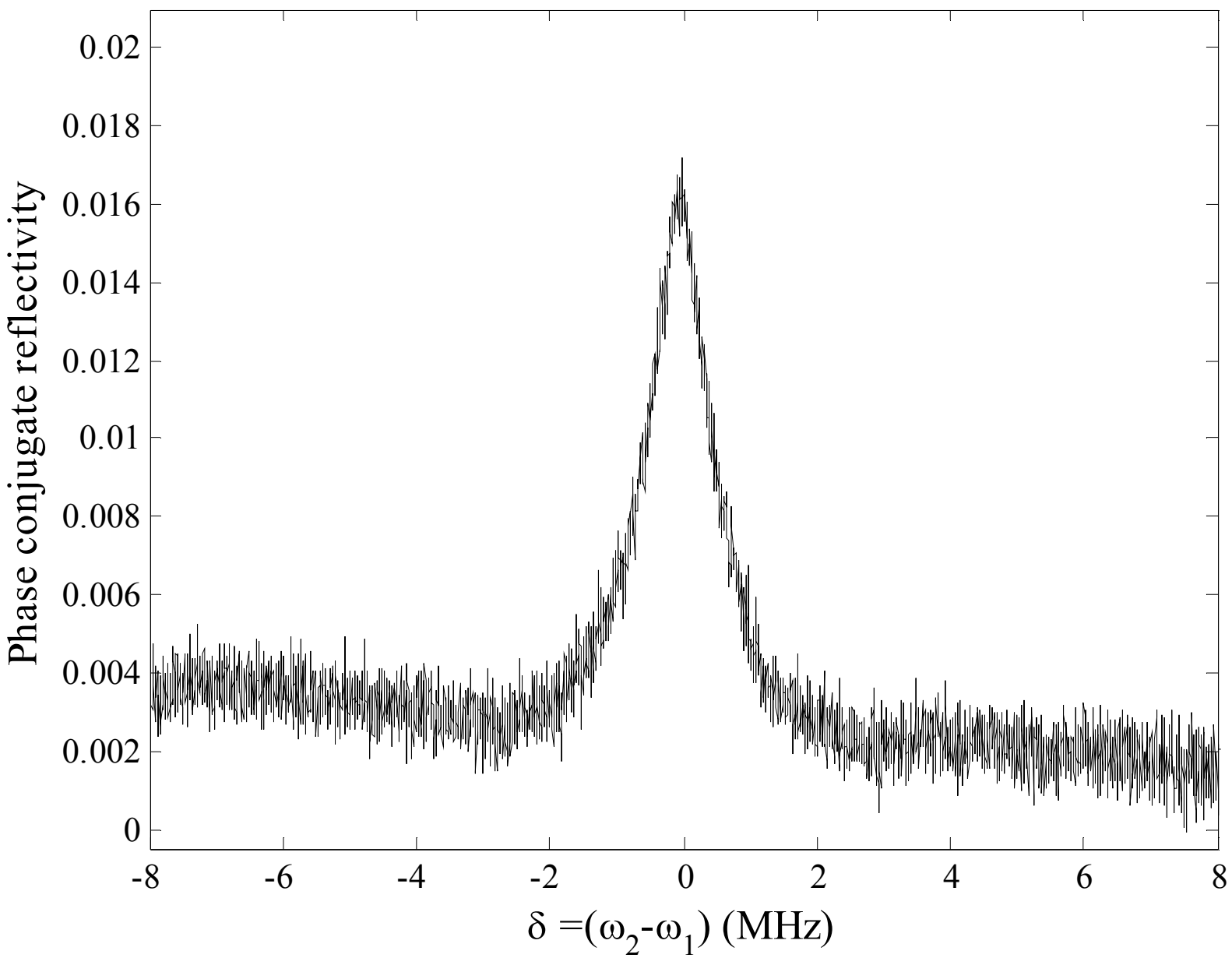

Fig 8 Optical phase conjugate signal produced by a back-propagating pump